\documentclass{moriond}

\usepackage{amstext}
\usepackage{xspace}

\def\Journal#1#2#3#4{{#1} {\bf #2}, #3 (#4)}

\def\PRL{\em Phys. Rev. Lett.}
\def\PRD{{\em Phys. Rev.} D}

\def\JHEP{JHEP}
\def\EPJ{Eur. Phys. J.}
\def\JINST{JINST}

\def\be{\begin{equation}}
\def\ee{\end{equation}}
\def\bea{\begin{eqnarray}}
\def\eea{\end{eqnarray}}

\newcommand{\Photo}{\includegraphics[height=35mm]{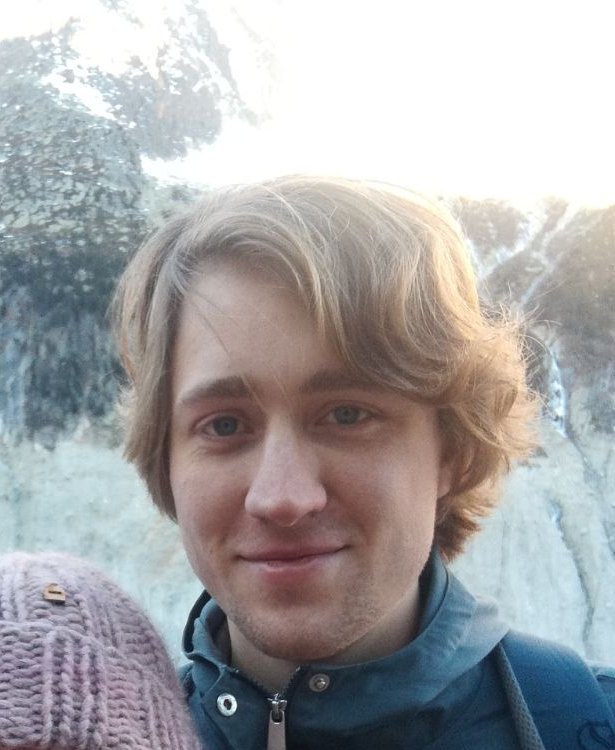}}

\newcommand{\fbinv}{$\text{fb}^{-1}$\xspace}
\newcommand{\tgq}{$\text{t}\gamma\text{q}$\xspace}
\newcommand{\tzq}{$\text{tZq}$\xspace}
\newcommand{\tttt}{$\text{t}\overline{\text{t}}\text{t}\overline{\text{t}}$\xspace}
\newcommand{\ttbar}{$\text{t}\overline{\text{t}}$\xspace}
\newcommand{\met}{$E_{\text{T}}^{\text{miss}}$\xspace}
\newcommand{\pt}{$p_{\text{T}}$\xspace}
\newcommand{\ttZ}{$\text{t}\overline{\text{t}}\text{Z}$\xspace}
\newcommand{\ttW}{$\text{t}\overline{\text{t}}\text{W}$\xspace}
\newcommand{\ttH}{$\text{t}\overline{\text{t}}\text{H}$\xspace}
\newcommand{\tWb}{$\text{tWb}$\xspace}

\newcommand\blfootnote[1]{%
  \begingroup
  \renewcommand\thefootnote{}\footnote{#1}%
  \addtocounter{footnote}{-1}%
  \endgroup
}

\begin{document}
\vspace*{4cm}
\title{Single top quark and rare top quark production at ATLAS and CMS}

\author{ Willem Verbeke }

\address{On behalf of the ATLAS and CMS Collaborations\\
Department of Physics and Astronomy,\\
Universiteit Gent, Gent, Belgium}

\maketitle\abstracts{
The latest results from ATLAS and CMS on single top quark production and rare production channels of top quarks at the LHC are presented.
}

\section{Introduction}
The dominant production mechanism of top quarks at the LHC is pair production through the strong interaction (\ttbar). Electroweak production of single top quarks is more rare, but provides essential complementary information for furthering our understanding of the SM. Being the heaviest particle in the SM, the top quark plays an important role in many new physics models. Measurements of rare processes involving top quarks are therefore essential tests of the SM.

The latest results on single top quark production and rare production channels of top quarks from ATLAS~\cite{ATLASDetector} and CMS~\cite{CMSDetector} are presented below. The majority of these results are based on data from proton-proton (pp) collisions at a center of mass energy of 13 TeV, and unless otherwise specified, these conditions are assumed. Analyses discussed in this document generally use events with electrons and muons in the final state, and therefore `leptons' will only refer to electrons and muons unless otherwise stated.
\blfootnote{2019 CERN for the benefit of the ATLAS and CMS Collaborations. CC-BY-4.0 license.}

\section{Single top quark production}
Production of single top quarks in the $t$-channel is highly sensitive to the proton's parton distribution functions (PDF) since the flavor of the incoming quark determines the charge of the top quark. Through this process one can additionally measure the CKM matrix element $V_{\text{tb}}$. Using events with a single isolated lepton and several jets in 35.9 \fbinv of pp collision data, CMS measured $\left|f_{\text{LV}} V_{\text{tb}}\right| = 1.00 \pm 0.08~\text{(exp)} \pm 0.02 ~\text{(theo)}$, where $f_{\text{LV}}$ is a modification factor, and the charge ratio $\sigma_{ t\text{-ch}, \text{t}} / \sigma_{ t\text{-ch}, \overline{\text{t}}} = 1.66 \pm 0.02 ~\text{(stat)} \pm 0.05 ~\text{(syst)}$.~\cite{Sirunyan:2018rlu} 

Combinations of all 7 TeV and 8 TeV measurements of single top quark production in the t- and s-channels and in association with a W boson, from ATLAS and CMS, were recently performed. These result in more precise measurements of each of the production cross sections, and the most precise direct determination of $V_{\text{tb}}$ to date. \cite{Aaboud:2019pkc} The combined value is $\left| f_{\text{LV}} V_{\text{tb}} \right| = 1.02 \pm 0.04 ~\text{(exp)} \pm 0.02 ~\text{(theo)}$, with a relative uncertainty of about 3.7\%, markedly smaller than in the best single measurement where the uncertainty is 4.7\%.~\cite{Aaboud:2017pdi}

The cross section and charge-ratio of $t$-channel single top quark production are measured differentially in the top quark transverse momentum (\pt) and rapidity, the charged lepton \pt and rapidity, and the \pt of the W boson from the top quark decay, by CMS using 35.9 \fbinv of data. The results are in agreement with predictions from several NLO event generators using multiple sets of PDFs. The cross section is additionally measured as a function of the top quark polarization angle, shown in Fig.~\ref{cmsfigures}, which can be used to probe the structure of the $\text{tbW}$ vertex. This is quantified by the so-called top quark spin asymmetry, which is found to be $A = 0 .439 \pm 0.032 ~\text{(stat + exp)} \pm 0.053 ~\text{(syst)}$, in excellent agreement with the SM value of $A = 0.436$ as predicted by \textsc{POWHEG}.~\cite{CMS-PAS-TOP-17-023}

Associated production of a single top quark with a W boson and b quark (\tWb) interferes with \ttbar because both processes have the same final state of two W bosons and two b quarks. The first probe of these interference effects is done by ATLAS in events with exactly two b jets and two leptons in 36.1 \fbinv of pp collision data. To attain sensitivity to the interference effects a differential cross section measurement of \ttbar + \tWb is done as a function of $m_{b\ell}^{\text{minimax}}$ $(= \text{min}\{\text{max}( m_{b_{1}\ell_{1}}, m_{b_{2}\ell_{2}} ) , \text{max}( m_{b_{1}\ell_{2}}, m_{b_{2}\ell_{1}} ) \})$, which is always smaller than $\sqrt{ m_{\text{t}}^{2} - m_{W}^{2} }$ for \ttbar at parton level. Different modeling schemes of the interference effects are then compared to the unfolded $m_{b\ell}^{\text{minimax}}$ distribution. The best modeling is found to be given by a \textsc{POWHEG-BOX-RES} simulation of the $\ell^{+}\ell^{-}\nu\overline{\nu} \text{b} \overline{\text{b}}$ final state which takes off-shell and interference effects into account, as shown in Fig.~\ref{atlasfigures}.~\cite{Aaboud:2018bir}

\section{Rare top quark production}
Production of a single top quark and a photon (\tgq) is sensitive to the top quark's charge, and its electric- and magnetic dipole moments. A search for this process is performed by CMS in events with a muon, a photon, and jets in 35.9 \fbinv of data. A boosted decision tree (BDT) is used to separate the signal from the background, and a fit to the BDT distribution results in a cross section measurement $\sigma(\text{pp} \to \text{t} \gamma \text{q}) \mathcal{B}(\text{t} \to \mu\nu\text{b}) = 115 \pm 17~ \text{(stat)} \pm 30~\text{(syst)} \; \text{fb}$, in a fiducial volume characterized by $p_{\text{T}}^{\gamma} > 25 \; \text{GeV}$, $\left| \eta^{\gamma} \right| < 1.44$, and $\Delta R (X, \gamma ) > 0.5$, where $X$ signifies a muon or parton. The signal has an observed (expected) significance of 4.4 (3.0) standard deviations (s.d.) over the background only hypothesis, the first evidence for \tgq.~\cite{Sirunyan:2018bsr}

\begin{figure}[h]
\centering
\includegraphics[width=.35\textwidth]{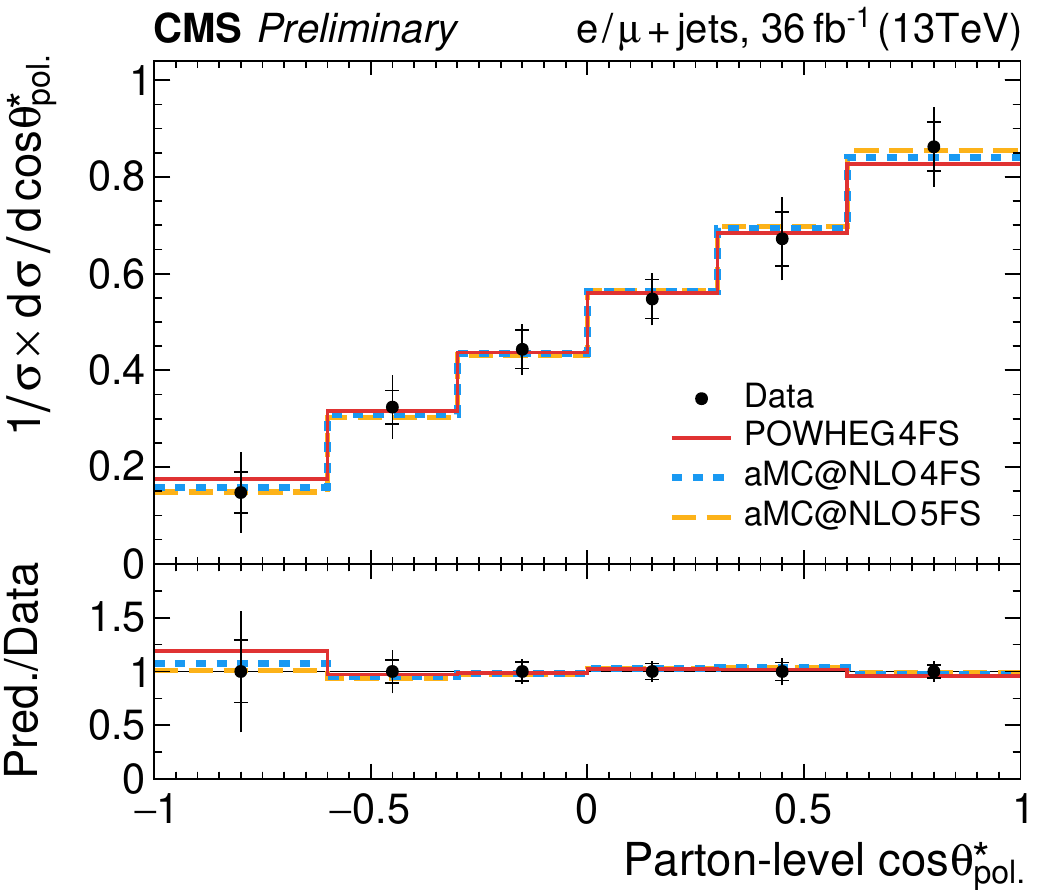}
\includegraphics[width=.24\textwidth]{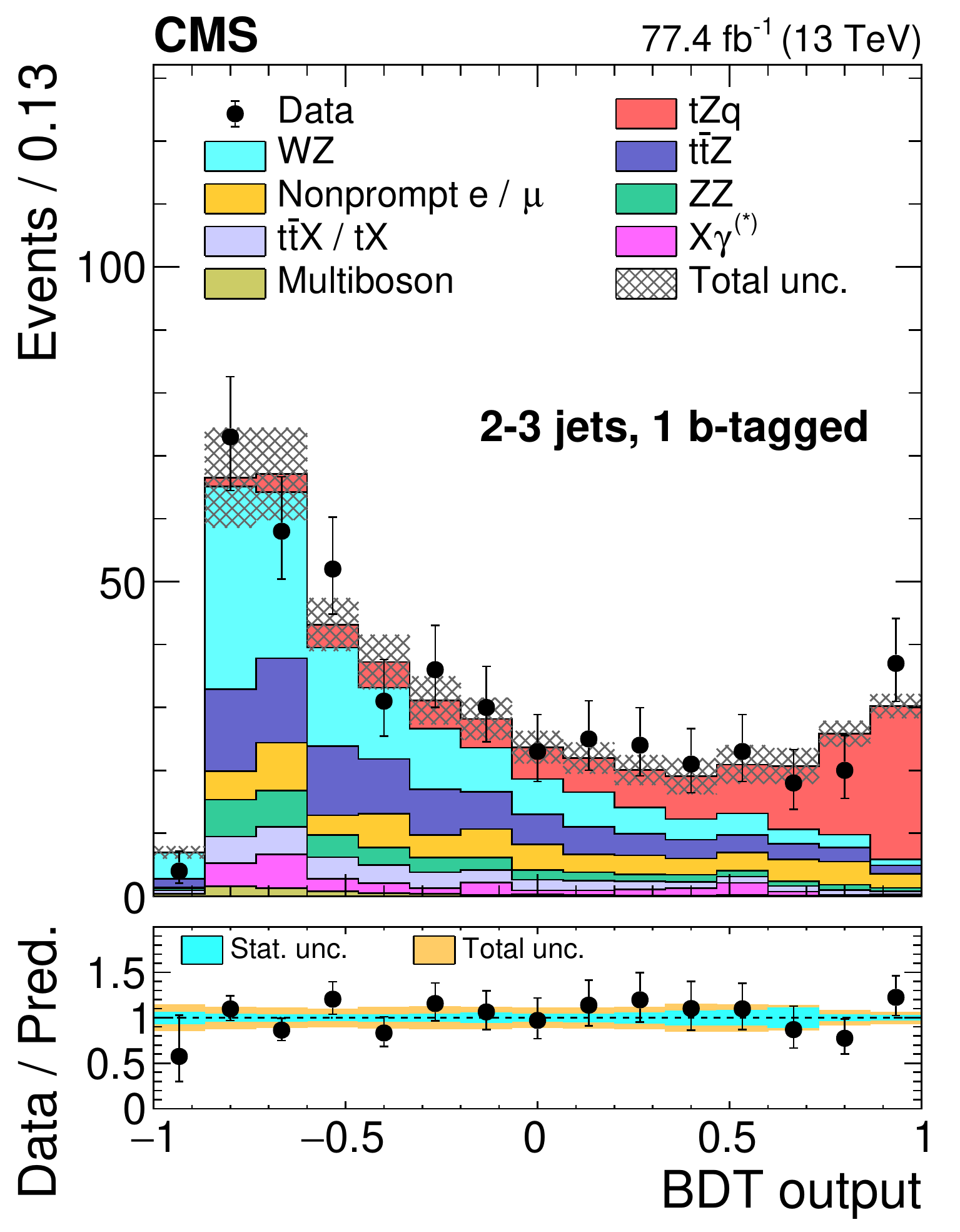}
\includegraphics[width=.39\textwidth]{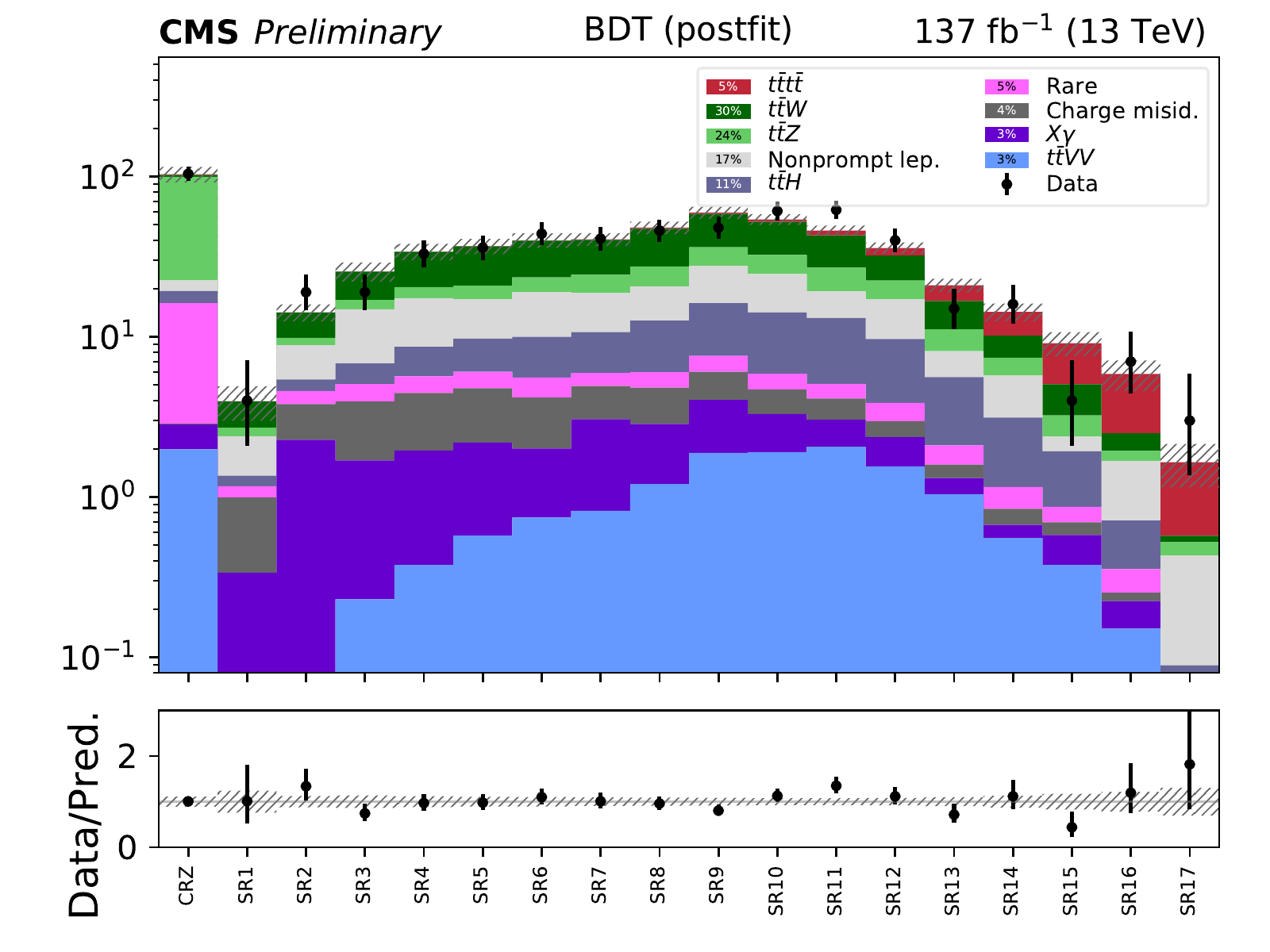}
\caption{Cross section of $t$-channel single top quark production as a function of the top quark polarization angle at parton level (left)~\protect\cite{CMS-PAS-TOP-17-023}, BDT distribution in the most important category of the \tzq search (middle)~\protect\cite{Sirunyan:2018zgs}, BDT distribution of the CMS multilepton \tttt search (right)~\protect\cite{CMS:2019nig}.}
\label{cmsfigures}
\end{figure}

Single top quark production in association with a Z boson (\tzq) is a rare process that until recently remained unobserved. It depends on the $\text{tZ}$ and $\text{WWZ}$ couplings, and is highly sensitive to the presence of new physics because of unitary cancellations in the SM. A new search for this process is done by CMS, using events with three leptons in 77.4 \fbinv of pp collision data. A crucial improvement is the usage of a BDT-based lepton identification, which increases the signal efficiency while lowering the background from nonprompt muons (electrons) by a factor 8 (2) compared to the lepton idenficiation used in the previous CMS search for \tzq. The \tzq cross section is extracted from a fit to BDT distributions in three event categories based on the number of jets and b jets. The result is $\sigma(\text{pp}\to \text{tZq} \to \text{t}\ell\ell\text{q}) = 111 \pm 13~\text{(stat)} ^{+11}_{-9}~\text{(syst)} \; \text{fb}$, for dilepton invariant masses above 30 GeV with $\ell = e, \mu, \tau$, in agreement with the SM expectation. The observed (expected) significance of the signal is 8.2 (7.7) s.d. above the background only hypothesis, marking the first observation of this process. The distribution of the BDT discriminant for events with 2 or 3 jets, one of them b-tagged, the most important event category, is shown in Fig.~\ref{cmsfigures}.~\cite{Sirunyan:2018zgs}

An extremely rare and unobserved process is the production of four top quarks (\tttt), which has a SM cross section $\sigma^{SM}( \text{t}\overline{\text{t}}\text{t}\overline{\text{t}} ) = 12.0 ~\text{fb}$. Because of the presence of four top quarks, the process is characterized by large jet and b jet multiplicities. About 40\% of \tttt events lead to a final state with one lepton or two leptons of opposite sign ($\ell/\ell^{+}\ell^{-}$ channel). In this final state the backgrounds, in particular that from \ttbar, are large. Decays to two leptons of the same sign, or three or more leptons (multilepton channel) have a branching fraction of only 12\%. Nonetheless such events are expected to form the most sensitive channel for \tttt searches because of the small backgrounds, which mainly come from \ttbar production in association with a W, Z or H boson. 

Using 36.1 \fbinv of data, ATLAS carried out searches for \tttt in both the $\ell/\ell^{+}\ell^{-}$ and multilepton channels. In the former channel, jets are reclustered using the anti-$k_{\text{T}}$ algorithm with a large distance parameter. These large-R jets are checked for compatibility with a hadronic top quark decay. Events are then categorized according to the number of jets, b jets and large-R jets, and events in each category are further binned in the amount of hadronic transverse energy. A unique feature of this analysis is the data-driven prediction of the \ttbar background, based on measurements of the probability of an additional jet in \ttbar events being b-tagged.~\cite{Aaboud:2018jsj} In the multilepton channel, a general search for new phenomena is carried out at high \met and high hadronic transverse energy. Events are binned according to the lepton flavor and the number of b jets, and the results are interpreted in terms of \tttt production.~\cite{Aaboud:2018xpj} The combination of the two aforementioned searches results in an observed (expected) signal significance of 2.8 (1.0) s.d. The observed deviation from the expected result is due to an upward fluctuation in the multilepton search.

Recently CMS also carried out a search for \tttt in $\ell/\ell^{+}\ell^{-}$ events with 35.9 \fbinv of data. Events are subdivided according to the number of leptons, their flavor, and the number of jets and b jets. A combination of BDT-based top-tagging and information about the event topology, hadronic activity and b-tagging is used to train several BDT discriminants, which are fit to obtain the final result.~\cite{CMS:2019cof} Combination with an earlier result in the multilepton channel~\cite{Sirunyan:2017roi} results in an observed (expected) signal significance of 1.1 (1.4) s.d.

The complete Run 2 dataset, corresponding to an integrated luminosity of 137 \fbinv, was used by CMS to carry out a new search for \tttt in the multilepton channel. Cut- and BDT-based analyses are performed simultaneously, with the former being a cross-check. The simulations of the most important backgrounds, \ttZ and \ttW, are corrected based on \ttbar data. A fit to the BDT distribution, shown in Fig.~\ref{cmsfigures}, results in a measured cross section $\sigma \left( \text{t} \overline{\text{t}} \text{t} \overline{\text{t}} \right) = 12.6^{+5.8}_{-5.2} ~\text{fb}$ and an observed (expected) signal significance of 2.6 (2.7) s.d. This is the most sensitive probe of \tttt production by a large margin at the time of writing. The cross section measurement can also be used to constrain the top quark yukawa coupling ($y_{\text{t}}$). One of the production diagrams of \tttt involves a virtual H exchange, so the cross section is proportional to $\left| y_{\text{t}} \right|^{4}$. The measured cross section, on the other hand, decreases as a function of $\left| y_{\text{t}} \right|$ as the expected \ttH background increases. The comparison of the measured and expected cross sections as a function of $\left|y_{\text{t}}/y_{\text{t}}^{\text{SM}} \right|$, results in a 95\% C.L. upper limit $\left|y_{\text{t}}/y_{\text{t}}^{\text{SM}} \right| < 1.7$.~\cite{CMS:2019nig}

\begin{figure}
\centering
\includegraphics[width=.35\textwidth]{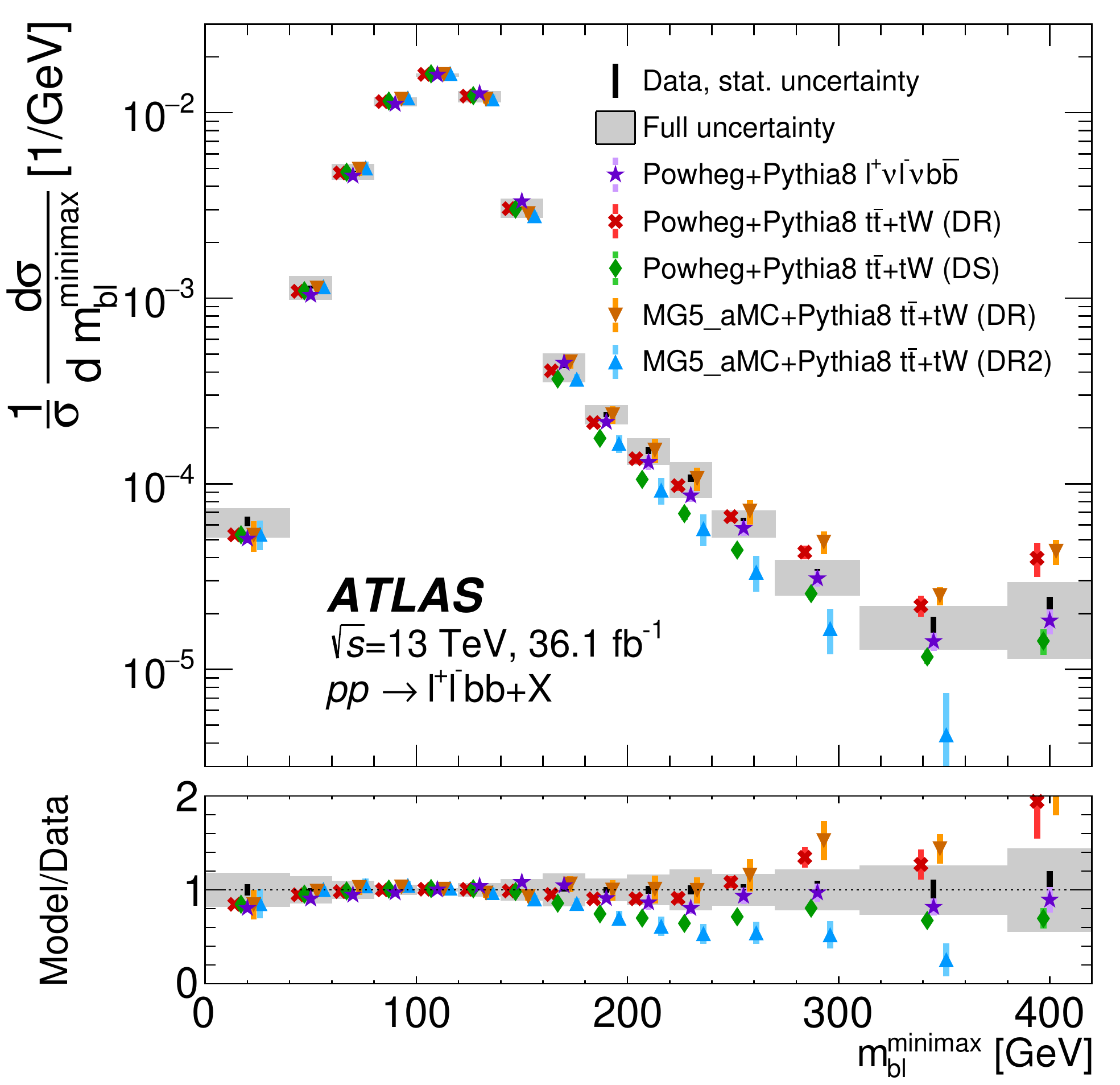}
\includegraphics[width=.3\textwidth]{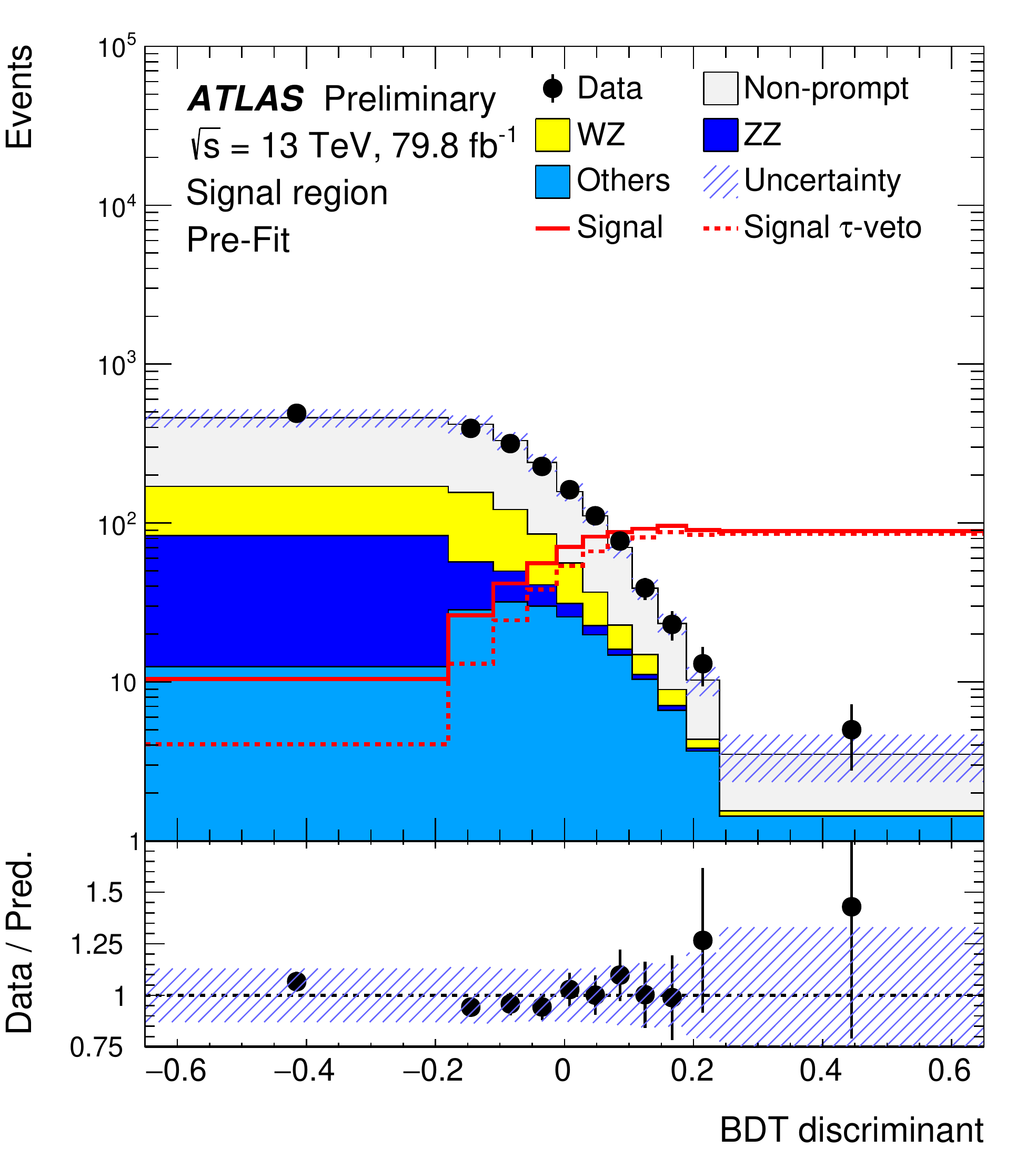}
\caption{Normalized cross section of \ttbar + \tWb as a function of $m_{b\ell}^{\text{minimax}}$ (left)~\protect\cite{Aaboud:2018bir}, BDT distribution of the cFLV search (right)~\protect\cite{ATLAS-CONF-2018-044}.}
\label{atlasfigures}
\end{figure}

Charged lepton flavor violation (cFLV) can occur in the SM by means of neutrino oscillations, but it is so rare that it is far beyond our current experimental reach. As such, the observation of cFLV would be a clear sign of the presence of new physics. The first direct search for cFLV in top quark decays is carried out by ATLAS in 79.8 \fbinv of pp collision data. The search targets the decay $\text{t} \to \ell^{\pm} \ell^{'\mp} \text{q}$, and uses events with three leptons and several jets out of which there is at most one that is b-tagged. A BDT discriminant, plotted in Fig.~\ref{atlasfigures}, is used to separate the signal from the background. The data are found to be consistent with the SM, and the resulting 95\% C.L. upper limits on the cFLV branching fraction are $\mathcal{B}(\text{t}\to \ell \ell^{'} \text{q}) < 1.86~(1.36^{+0.61}_{-0.37}) \cdot 10^{-5}$ if $\ell = e, \mu, \tau$, and $\mathcal{B}(\text{t}\to e \mu \text{q}) < 6.6~(4.8^{+2.1}_{-1.4}) \cdot 10^{-6}$ if $\ell = e, \mu$. This represents an improvement of several orders of magnitude compared to the previous best limits from indirect searches which are of the order of $10^{-3}$.~\cite{ATLAS-CONF-2018-044}

\section*{References}

\begin{thebibliography}{99}
\bibitem{ATLASDetector} ATLAS Collaboration, \Journal{\JINST}{3}{S08003}{2008}
\bibitem{CMSDetector} CMS Collaboration, \Journal{\JINST}{3}{S08004}{2008}
\bibitem{Sirunyan:2018rlu} CMS Collaboration, arXiv:1812.10514
\bibitem{Aaboud:2019pkc} ATLAS and CMS Collaborations, arXiv:1902.07158
\bibitem{Aaboud:2017pdi} ATLAS Collaboration, \Journal{\EPJ}{C77}{2017}{531}
\bibitem{CMS-PAS-TOP-17-023} CMS Collaboration, CMS-PAS-TOP-17-023 (2019), https://cds.cern.ch/record/2667180
\bibitem{Aaboud:2018bir} ATLAS Collaboration, \Journal{\PRL}{121}{152002}{2018}
\bibitem{Sirunyan:2018bsr} CMS Collaboration, \Journal{\PRL}{121}{221802}{2018}
\bibitem{Sirunyan:2018zgs} CMS Collaboration, \Journal{\PRL}{122}{132003}{2019}
\bibitem{Aaboud:2018jsj} ATLAS Collaboration, \Journal{\PRD}{99}{052009}{2019}
\bibitem{Aaboud:2018xpj} ATLAS Collaboration, \Journal{\JHEP}{1812}{039}{2018}
\bibitem{CMS:2019cof} CMS Collaboration, CMS-PAS-TOP-17-019 (2019), https://cds.cern.ch/record/2666712
\bibitem{Sirunyan:2017roi} CMS Collaboration, \Journal{\EPJ}{C78}{140}{2018}
\bibitem{CMS:2019nig}CMS Collaboration, CMS-PAS-TOP-18-003 (2019), https://cds.cern.ch/record/2668710
\bibitem{ATLAS-CONF-2018-044} ATLAS Collaboration, ATLAS-CONF-2018-044 (2018),\\ https://cds.cern.ch/record/2638305

\end{thebibliography}
\bibstyle{unsrt}

\end{document}